# Spintronics on Chiral Objects

See-Hun Yang

IBM Research – Almaden, San Jose, CA 95120, USA

e-mail) seeyang@us.ibm.com

**Abstract**

Chirality, handedness, is one of the most fundamental intriguing asymmetries in nature. By definition, chiral objects cannot be superimposed onto each other after mirror reflection operation. Numerous examples of chiral structures can be found in nature, for example, chiral molecules and chiral magnetic nanostructures. Moving electrons are associated with handedness by their own spins due to spin-orbit interaction thus exhibiting various emergent phenomena as they interact with chiral materials, which otherwise would not be observed in achiral systems. This new paradigm allows the potential development of new forms of devices or methods by utilizing reciprocal interaction of chiral objects with moving electron spins. This review updates the remarkable progresses in Spintronics on Chiral Objects that have been made over the past few years providing provides an outlook for new opportunities and potential applications with new insights.

Since spintronics was born in the wake of the discovery of giant magnetoresistance (GMR) from spin valves, it has made remarkable progress achieving a few seminal milestones such as oscillatory interlayer exchange coupling[1], spin-transfer torque (STT)[2], large tunneling magnetoresistance (TMR) from crystalline MgO barrier based magnetic tunnel junctions (MTJs)[3,4], and spin-orbit torque (SOT)[5–7] to date. These discoveries have revolutionized the information technology changing our daily lives with the massive applications, e.g., read-head in hard-disk drive and STT magnetic random access memory (STT-MRAM) over the past three decades. The spintronics utilizes the reciprocal coupling of moving electron spins to magnetic elements in magnetic structures. This is possible since the spins are polarized by spin-dependent scattering at the magnetic moments and reversely the spin angular momenta are transferred to the moments inducing spin current and STT, respectively. Note that spin currents can also be induced by spin-orbit interaction (SOI), e.g., via spin Hall effect (SHE)[8] or Rashba effect[9], that does not need magnetic moments to scatter by the spins. The SOI-induced spin current has recently emerged as a promising source of torques to manipulate neighboring magnetic nanostructures efficiently that can serve the development of non-volatile memory, logic and neuromorphic devices thus opening the door to spin-orbitronics about ten years ago.

In the meantime, another means to induce spin currents has garnered enormous attention in interdisciplinary communities. The new knob coupled to spins to generate spin current is one of the most intriguing asymmetries in nature, *Chirality*. Chirality corresponds to handedness such that chiral objects with different handedness cannot be superimposed over each other while each has the mirror-reflected other. Numerous chiral objects can be found in nature. One of the archetypal examples is chiral molecules, i.e., enantiomers in stereoisomers (3D isomers). Note that moving spins are also chiral due to their mirror-reflection asymmetry depending on relative momentum and spin orientations (Fig. 1a). This allows chiral objects to couple to the moving spins. Consequently, spins with a certain handedness can travel through chiral molecules with the same handedness in a longer distance, and vice versa, needing no magnetic moments here, leading to spin-filtering and spin current, namely chirality-induced spin selectivity (CISS)[10]. Importantly, the CISS does not need a large SOI either. Although there has yet been a complete consensus on the detailed mechanism of spin coupling to chirality in chiral molecules, it is believed that the combination of a dipole electric field with exchange interaction may give rise to large spin and chirality dependent transmission/tunneling[11–14]. In early years, while the observed electron spin filtering in chiral molecules was much smaller than theoretical predictions[15], a significant magnetochiral anisotropy in photoluminescence intensity was reported from the experiment in illumination of unpolarized photons on chiral molecules in the presence of magnetic fields[16].

Later in early 2010s large spin filtering started to be observed from the injection of unpolarized electron spins into chiral DNA[17,18] or other molecules[19,20] in which the spin polarization was found to be linearly proportional to the length of DNA molecules[17] (Fig. 1b). These findings have been led to the development of the CISS-based spin valves that have non-magnetic metallic layers in the bottom followed by chiral molecules that are coated with a thin oxide layer/ferromagnetic layer. Note here that the chiral molecules play a role as spin polarizer while ferromagnetic layer acts like spin analyser thus displaying GMR or TMR-like behaviour as an external magnetic field switches the ferromagnetic layer even at room temperature. When an opposite handedness in chiral molecules is used, i.e, chirality switches, the sign of magnetoresistance is observed to switch as expected from the CISS[21,22]. Sizeable magnetoresistance reflecting high-spin selectivity have been reported for wide range of chiral molecules. Remarkably, it was recently reported that light irradiation or thermal treatment can modify the structure of artificial chiral molecules thus switching the handedness of chiral molecules and consequently spin-polarization, which allows the development of reconfigurable spin valve based memory or logic[23].

The CISS can accumulate spins in non-magnetic layers that behaves like magnets in which the accumulated spins can be detected by Hall devices[24]. For example, when the Hall bar that is made of two-dimensional electron gas (2DEG) covered by chiral molecules/nanoparticle assembly is exposed to a circularly polarized light, the measured Hall voltage reveals striking contrasts depending on both the handedness of chiral molecules and the sign of circular polarization of the light. This shows that the measured Hall signal is anomalous Hall-like from the accumulated spins by CISS spin current in magnetic 2DEG layer. Moreover, as the CISS-induced spin current is large enough, magnetization of a magnetic element can be switched by the STT induced by the spin current. This was demonstrated by the application of chiral molecules on perpendicularly magnetized ferromagnetic layer. The switching directions are shown to be consistent with the signs of handedness of chiral molecules[25]. In addition, magnetic switching by CISS is found to be efficient and scalable allowing the development of low-power high-density devices[25]. It was recently shown that the adhesion rate of enantiomer chiral molecules on perpendicularly magnetized layer highly depends on the handedness of the enantiomer and magnetization direction of magnetic layer thus having the potential application to biochemistry and pharmaceutical industries[26]. Yet, along with these advantages, there are still a few challenges, e.g. slow switching of chirality in chiral molecules and insufficient understanding of CISS mechanism to further enhance spin selectivity.

Another emerging class of chiral objects for spintronics is chiral magnetic nanostructures such as chiral magnetic domain walls (DWs) and chiral magnetic skyrmions. These chiral magnetic objects display unique intriguing interactions with moving spins that otherwise does not exist in achiral counterparts. Handedness of chiral domain walls and chiral skyrmions are typically set up by Dzyaloshinsky-Moriya exchange interaction (DMI) that are mostly formed by broken inversion symmetry like interfaces or noncentrosymmetric crystallinity. Interface DMI (iDMI) favors Néel-type DWs and skyrmions in perpendicularly magnetized thin films[27,28] (Fig. 1c) while bulk DMI stabilizes Bloch-type DWs although Néel-type chiral DWs are observed from ultrathin centrosymmetric Heusler films[29]. Combination of this chiral nature with SOT gives rise to very efficient current induced motion of chiral magnetic objects compared to conventional STT driven achiral counterparts[30,31]. For example, the damping-like SOT rotates the magnetizations away from Néel direction thus inducing out-of-plane oriented DMI field driven torques. Consequently, chiral DWs and skyrmions move along current flow direction irrespective of DW configurations or skyrmion polarities when the signs of SHE and DMI are either all positive or negative. This symmetry in current driven motion of chiral DWs can be lifted by external conditions such as in-plane fields along the iDMI field axis[30,32] or geometrical effects like branches[33] or curves[34]. For example, the DW velocity increases with the increasing field along the iDMI field direction strengthening the chirality, while the velocity decreases with the fields that compensates the iDMI field weakening the chirality (Fig. 1c). Especially, the geometric effect on the motion of chiral objects results from the DW tilting[35,36] caused by the interplay of SOT, DMI and magnetostatic energies thus forming the potential challenges to development of high-density devices. Composite chiral magnetic objects can overcome these challenges as will be discussed below[33,34,37]. On the other hand, conventional damping-like STT moves the chiral magnetic objects similarly to achiral counterparts except the reduced velocity. The velocity reduction is due to the suppression of precession by the DMI field at a given current density. When a field is applied along the DMI field axis, the velocity-field curve forms a dome-like shape[38], the center of which sits at the negative value of DMI field[29].

As two chiral magnetic objects are coupled antiferromagnetically forming a composite structure in namely synthetic antiferromagnets (SAFs), not only many challenges described above can be overcome but additional benefits are followed[39,40]. For example, the composite chiral objects (a) move more efficiently by current due to powerful exchange coupling torque (ECT)[40], (b) are more robust and thermally stable, (c) are inert to fields/geometric effects[33,34], and (d) have lower threshold current density above which the objects start to be depinned[39,41]. Note here that the ECT is effective only for antiferromagnetic coupling not for ferromagnetic coupling. Moreover, the torque strength increases as the net magnetization becomes more compensated, which can be extended to ferrimagnets[42–45] and antiferromagnets[46,47] to increase the DW velocity even further, all of which have the potential to develop high-density fast chips[37].

Recently, as the disparity between two torque strengths exerted on component chiral DWs exceeds a threshold value in a weakly coupled SAF, a new current-driven DW dynamic phase was observed thus showing dramatic reduction of DW velocity accompanied by synchronous oscillation/precession of DW magnetizations due to the angular momentum conservation via exchange coupling and angular momentum transfer torque, namely chiral exchange drag anomaly[48]. This finding has the potential to be used as oscillators to generate radio frequency currents. Neuromorphic functionality of chiral DW based devices have been recently paid lots of attention due to threshold behaviors and analogue features thus providing vast opportunities for application to artificial intelligence[49–51]. There are several challenges for commercialization of spintronics on chiral DWs such as further reduction of threshold current density while obtaining decent thermal stability. Enhancing SOT by engineering of interfaces/textures or exploring new materials would be one of the promising routes to overcome these challenges. Recent studies on laterally coupled chiral DWs show the potential to be used for various applications in SOT devices, skyrmion devices, and memory/logics[52,53].

Chiral magnetic skyrmions, 2D chiral magnetic objects in contrast with 1D counterpart DWs, have been highlighted due to the potential usage as data bits when they can be individually isolated. Stabilization of skyrmions at room temperature and zero field has been pursued by utilizing exchange field[52], field history[54], exchange-bias field[55], multi-stacking of films[56], and synthetic antiferromagnets[57]. Chiral skyrmions have additional topological nature compared to chiral DWs that gives rise to the traverse force to current direction, the sign of which is determined by skyrmion number, namely skyrmion Hall effect[58]. The skyrmion Hall effect is typically parasitic against the technological use but may be removed using ferrimagnets[59] and synthetic antiferromagnetic wire[60]. As a reciprocal topological property to skyrmion Hall effect, skyrmions apply traverse force to the moving spins giving rise to anomalous velocity perpendicular to the current flow direction, namely topological Hall effect (THE)[61,62] that has the potential to be used as readouts in skyrmion-based devices or artificially fabricated chiral magnetic structures like artificial chiral spin ice[63–65].

Large anomalous Hall (AHE) and Nernst effects that used to be observed from ferromagnets only[66,67] have been recently observed from chiral non-collinear antiferromagnets[68] such as $Mn_3Sn$[69], $Mn_3Ir$[61], and $Mn_3Ge$[71] due to non-vanishing Berry phase. These exotic materials exhibit small coercivity allowing the large magneto-transport coefficients to be used as readouts in chiral antiferromagnets-based devices. In addition, these materials can serve as an excellent spin current source since SHE shares the same mechanism with AHE[72]. Indeed, a large spin current and an exotic magnetic SHE were recently reported from non-collinear chiral antiferromagnets $Mn_3Ir$[70] and $Mn_3Sn$[73], respectively, thus constituting potential applications to three-terminal SOT spintronic devices. In addition, Chiral DWs in antiferromagnets can be used as a source of terahertz excitation of spin waves[46,74] showing the potential application to development of terahertz oscillator.

Spintronics based on chiral magnetic nanostructures poises not only to be used for various applications in memory, logic, and large-scale storage but to allow development of new forms of devices such as memristic analogue and neuromorphic devices. For example, chirality is expected to amplify spin currents further than those that are induced by SOI-only so that more efficient three terminal memory is possible. Although there are some roadblocks for practical applications and commercialization, several approaches are promising to overcome these as described above. In summary, the future of spintronics on chiral objects is bright showing that the useful and powerful chips based on the new combination of spins with chirality will reach the market soon.

**Figure Captions**

**Fig. 1**. Schematics of chiral nature of (a) moving spins, (b) chiral DNA molecules, and (c) chiral magnetic domain walls. The middle lines correspond to mirrors. (a) A spin $\vec{s}$ moves with a velocity $\vec{v}$. The spin orientation is coupled to the velocity direction in two different ways, right- and left-handedness. (b) Chiral DNA molecules twist around in right- and left-handed way. (c) Magnetizations $\vec{M}$ in Néel-type domain walls that are out-of-wall-plane twist around in right- and left-handed way. The Néel wall is stabilized by local DMI fields $\vec{H}_{DM}$.

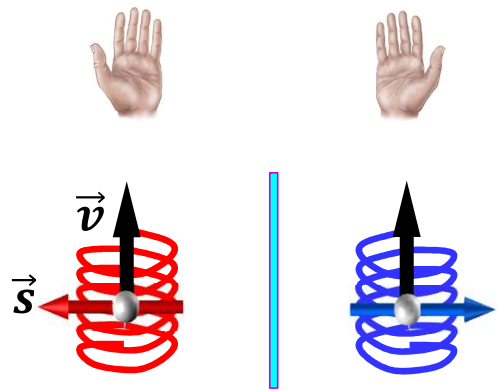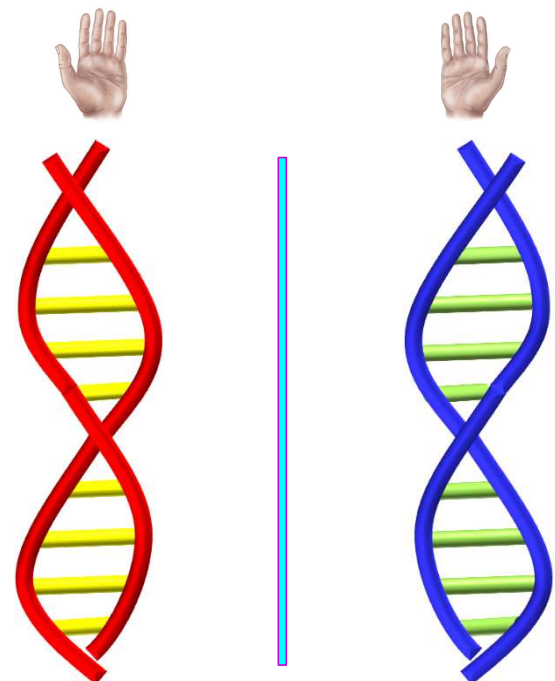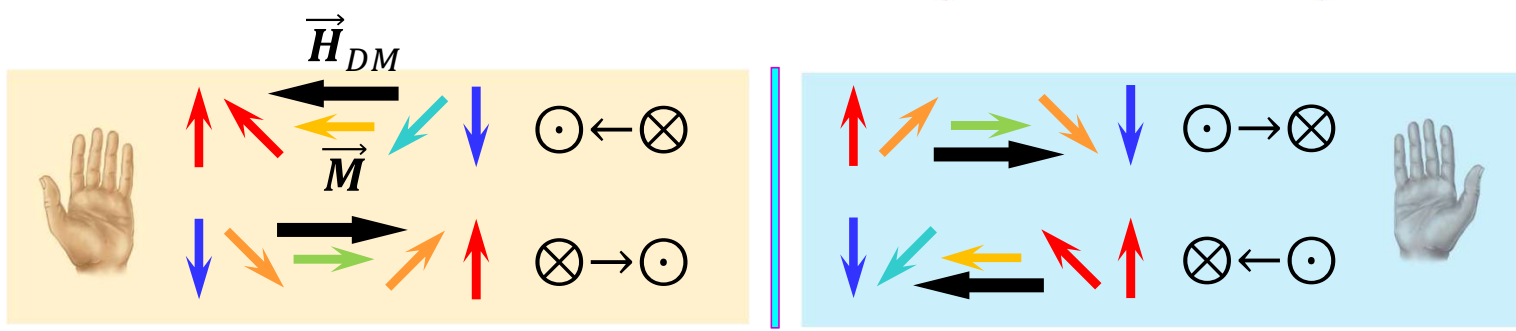

Fig. 1